\documentclass[twocolumn, secnumaRabic, amssymb, nobibnotes, aps, showpacs,superscriptaddress,longbibliography]{revtex4-1}%,
\usepackage{graphicx,subfigure,amsmath}%
\usepackage{color}
\usepackage[colorlinks=true, linkcolor=blue,urlcolor=blue,anchorcolor=blue,citecolor=blue,bookmarksnumbered]{hyperref}

\usepackage{amsmath,amstext}
\usepackage{graphicx}
\usepackage{booktabs}
\usepackage{multirow}
\definecolor{green}{rgb}{0.8,0.98,0.83}
\usepackage{float}
\usepackage{ulem}
\usepackage{braket}
\usepackage[ruled,linesnumbered,norelsize]{algorithm2e} 
\usepackage{amsmath,amssymb}
\usepackage[T1]{fontenc}
\begin{document}

\title{Reinforcement-Learned Electric-Field Sensing with Asymmetrically Blockaded Rydberg Arrays}
	\author{Shi-Qiang Qiao}
	\affiliation{School of Physics, Anhui University, Hefei 230601,  China}
	\author{Xue-Ke Song}\email[]{songxk@ahu.edu.cn}
	\affiliation{School of Physics, Anhui University, Hefei 230601,  China}
	\author{Shi-Lei Su}\email[]{slsu@zzu.edu.cn}
	\affiliation{Quantum Information Institute, School of Physics and Laboratory of Zhongyuan Light,
		Zhengzhou University, Zhengzhou 450001, China}
	\affiliation{Institute of Quantum Materials and Physics, Henan Academy of Science, Henan 450046,  China}
	\author{Liu Ye}
	\affiliation{School of Physics, Anhui University, Hefei 230601,  China}
	\author{Dong Wang}
	\affiliation{School of Physics, Anhui University, Hefei 230601,  China}
	
	\date{\today}

\begin{abstract}
We present a reinforcement learning-optimized Rydberg electrometer based on the asymmetric blockade effect and achieve high-sensitivity electric field sensing in Rydberg arrays. Microwave dressing induces asymmetric blockade to suppress interactions between target atoms, while keeping the coupling between the central control atom and target atoms field-tunable near Förster resonance. The field-regulated blockade radius affects the detectable atomic population signals, thereby enabling electric field sensing via state-selective readout.
In planar atomic arrays, classical Fisher information exhibits near-quadratic scaling with atom number and approaches the Heisenberg limit. Reinforcement learning-designed composite pulses greatly enhance quantum Fisher information by up to one order of magnitude compared with single $\pi$ pulses. We further establish a compact six-atom spherical configuration for vector electrometry, in which field orientation is extracted from calibrated axial populations, and weak bias fields eliminate dipole-dipole-induced sign and magic-angle ambiguities.
Numerical tests against Rabi frequency deviation, positional error, residual inter-target coupling and projection noise demonstrate the reliability of this scheme. This work provides an experimentally viable approach to realize high-precision three-dimensional Rydberg electric field sensing.
\end{abstract}
	\maketitle
	\section{Introduction}
	
	Precise measurement of electric fields is crucial in a wide range of applications, from fundamental physics experiments and astrophysical observations to industrial monitoring and biomedical imaging \cite{hitchcock2004radio}. Classical microwave field measurement relies on electric dipole antennas, which transduce incident radiation into measurable currents via electronic circuits. Although well-established, such systems are ultimately limited by electronic thermal noise and lack microscopic spatial resolution, motivating the pursuit of quantum-enhanced alternatives \cite{3601,1589516}. Quantum sensors, which exploit quantum coherence and entanglement, promise to overcome these classical limits by harnessing the exquisite sensitivity of quantum states to external perturbations \cite{RevModPhys.89.035002,RevModPhys.82.1155,https://doi.org/10.1002/1521-3978(200009)48:9/11<771::AID-PROP771>3.0.CO;2-E,6gql-zgkb,PhysRevLett.123.230401,Montenegro2025,PhysRevA.100.032104}. Among various platforms, Rydberg atoms have emerged as a particularly promising candidate for electric-field sensing due to their large polarizability, strong and tunable dipole-dipole interactions, and long-range Rydberg blockade effect \cite{PhysRevResearch.2.043130,Adams_2020,PhysRevA.102.062410,Fan_2015,kitson2025sensingelectricfieldsrydberg,6910267,Wu2025,Anand2024,PhysRevA.109.012619}. The Rydberg blockade—where the excitation of one atom to a Rydberg state suppresses the excitation of nearby atoms—provides a mechanism to create highly correlated many-body states that can encode environmental parameters with high fidelity \cite{PhysRevLett.85.2208,PhysRevLett.87.037901,gallagher1994rydberg,Wu:2023axc,Urban2009,PhysRevApplied.19.044007}.
	
	The sensing capability of Rydberg atoms can be dramatically enhanced near a Förster resonance, where the energy defect between a pair of Rydberg states is tuned to zero by an external electric field \cite{PhysRevA.65.063404,PhysRevX.2.031011}. At this resonance, the interatomic interaction undergoes a crossover from a van der Waals (\(R^{-6}\)) to a resonant dipole-dipole (\(R^{-3}\)) regime, leading to a significant enlargement of the Rydberg blockade radius \cite{walker2012entanglement,PhysRevA.109.022613,PRXQuantum.4.020335}. This strong field-dependence of the blockade radius forms the physical basis for high-sensitivity electrometry, as demonstrated in proposals and experiments using Rydberg atomic system\cite{wade2017real,facon2016sensitive,Schlossberger2024,HanYuLong2026,Ding2022,PhysRevB.111.144313}. Although the traditional four-level EIT-AT scheme, after optimization through techniques such as superheterodyne, can approach the standard quantum limit (SQL)~\cite{jing2020atomic,10.1063/5.0069195,PhysRevLett.117.113601,su2026broadbandheterodynemicrowavedetection}, theoretical analysis indicates that it is limited by absorption loss and cannot truly break through the SQL limit~\cite{PhysRevA.104.043103}. To surpass the classical limit, a quantum enhancement mechanism needs to be introduced, such as a single Rydberg atom electric dipole based on the Schrödinger cat state, multi-body critical enhancement sensing, or compressed optical spectroscopy \cite{ding2022enhanced,Prajapati:21,Yan2025,Wang2026}. Moreover, the traditional schemes mostly rely on fixed pulse sequences and empirical parameter tuning, lacking systematic optimization capabilities, and mainly focus on the detection of scalar fields, while the complete reconstruction of vector electric fields (a more general and informative quantity) has basically not been explored. These control and functional limitations restrict its development towards higher precision and multi-dimensional information sensing.
	
	The latest advancements in the fields of quantum control and machine learning offer new approaches to overcome these limitations. In the area of quantum control, the neutral atom array technology based on optical tweezers enables the realization of programmable Rydberg multi-body arrays, providing an ideal platform for constructing quantum sensors with controllable spatial structure and scalable atomic numbers \cite{Pause:24,PhysRevLett.130.180601,desantis2026realizationcavitycoupledrydbergarray,PhysRevLett.132.113601}. Simultaneously, reinforcement learning (RL), as a powerful optimization tool, has been successfully applied to design optimized pulse sequences for rapid state preparation and parameter estimation, achieving Heisenberg scaling of sensing enhancement in terms of resource dimensions such as pulse depth and atomic numbers \cite{PhysRevLett.134.120803,5591-xjfr,PhysRevA.109.062609,Porotti2023,PhysRevLett.121.150503}. Integrating RL-optimized pulse control with  optical-tweezer-trapped Rydberg arrays allows for the full exploitation of synergistic spatial and temporal correlations. This approach paves the way for a new generation of electric-field sensors characterized by both high sensitivity and exceptional spatial resolution.
	
This work combines three individually established but rarely integrated ingredients: microwave dressing for asymmetric blockade, F\"{o}rster-enhanced control-target interactions for sharp field-dependent population response, and reinforcement learning for optimizing composite pulse phases. In this protocol, the blockade radius dependent on the electric field is converted into a directly measurable population signal, and is further optimized through composite phase control. Based on this combination, we obtain the following results. First, for a planar array, the full-excitation population measurement yields finite-size near-quadratic scaling of classical Fisher information with atom number ($F_C \propto N^b$ with $b \simeq 1.88$--$2.09$ in the simulated range), while the corresponding quantum Fisher information scales superlinearly. Second, RL-optimized composite pulses further improve the final-state quantum Fisher information within the simulated pulse-depth range, with fitted exponents $1.90$--$2.44$. Third, for a spherical array with six target atoms, the field direction is inferred from calibrated axial populations, and a weak bias field resolves the sign and magic-angle ambiguities inherent to the $1-3\cos^2\theta$ dipole-dipole response, thereby enabling full vector electrometry. Fourth, robustness tests against Rabi-frequency errors, position errors, residual target-target coupling, and projection noise confirm that the protocol maintains high performance under realistic experimental imperfections. Collectively, these results establish an experimentally accessible pathway toward quantum-enhanced electric-field sensing with both high precision and full directional resolution, bridging the gap between many-body quantum control and practical Rydberg electrometry.

The paper is organized as follows. Section~\ref{sec2} introduces the asymmetric blockade model and the effective Hamiltonian of the Rydberg array system. Section~\ref{sec3} presents the planar-array electric-field sensing, including the field-dependent population response and the finite-size scaling of classical and quantum Fisher information. Section~\ref{sec:rl} demonstrates the reinforcement-learning-optimized composite pulse sequences that further enhance the quantum Fisher information. Section~\ref{sec4} extends the scheme to three-dimensional vector electrometry using a minimal spherical array, where a weak bias field resolves the angular ambiguities. Section~\ref{sec:implementation} discusses implementation considerations, including calibration requirements, robustness against experimental imperfections, and signal-to-noise ratio under projection noise. Section~\ref{sec5} concludes the manuscript.
	\section{MODEL AND Hamiltonian}\label{sec2}
	
\begin{figure}
	\centering
	\includegraphics[width=8.6cm,height=4cm]{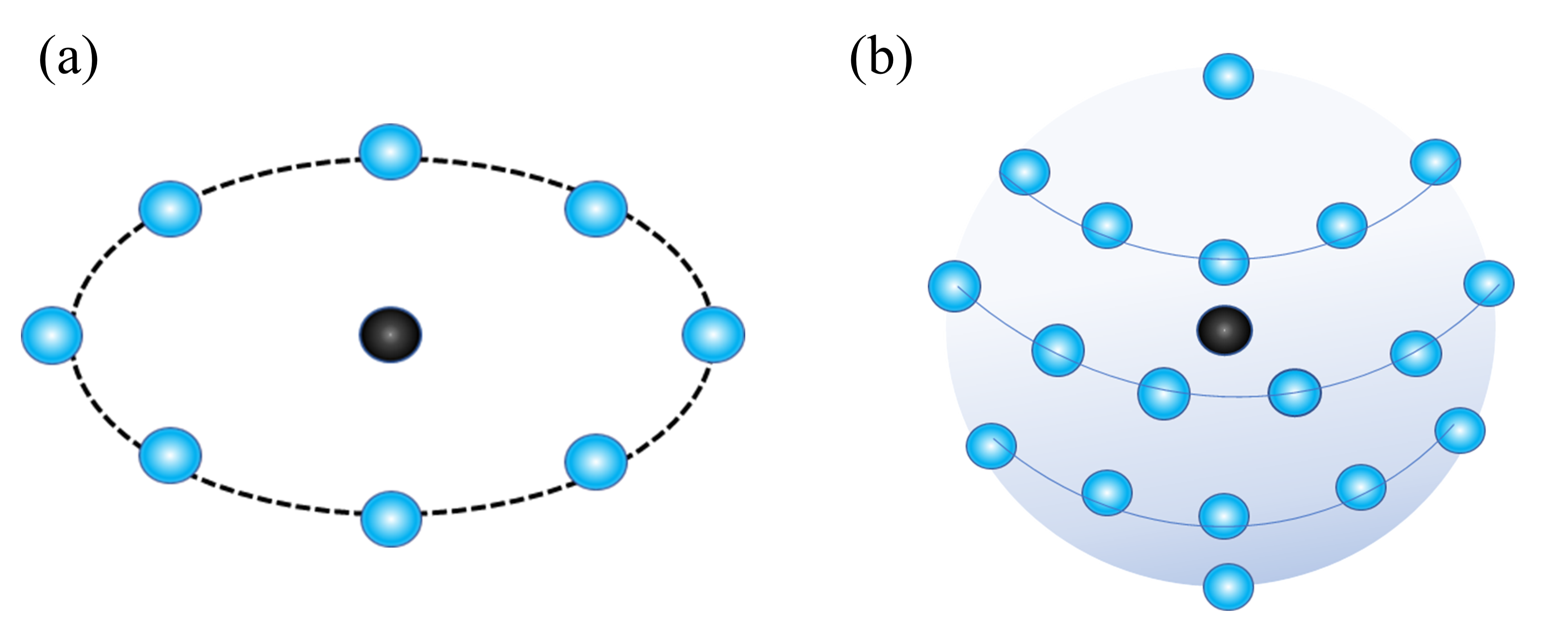}
	\caption{Sensor geometries considered in this work. A central control atom (black) is surrounded by target atoms (blue). (a) Planar circular array used for scalar electric-field sensing. (b) Spherical array used for vector sensing. Microwave dressing is designed to suppress target-target interactions while retaining field-sensitive control-target interactions.}
	\label{fig1}
\end{figure}

\begin{figure*}[t]
	\centering
	\includegraphics[width=0.86\textwidth]{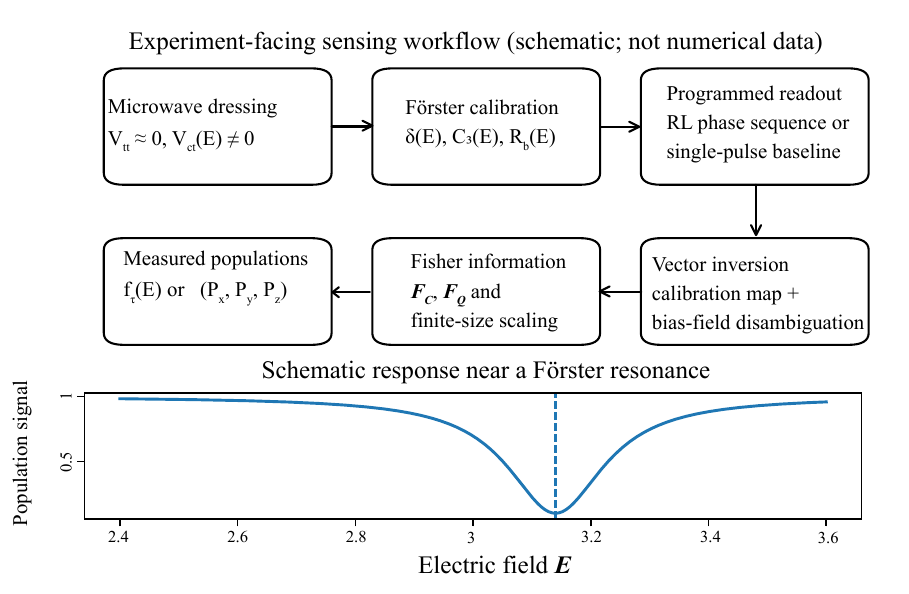}
	\caption{Experiment-facing workflow of the proposed protocol. Microwave dressing first prepares an asymmetric blockade configuration, the F\"{o}rster response is calibrated through $\delta(E)$ and $C_3(E)$, and the final population readout is performed either with a single-pulse baseline or with an RL-optimized phase sequence. The lower curve is a qualitative schematic of a population dip near resonance and is not numerical data.}
	\label{fig:workflow}
\end{figure*}

We consider the two geometries shown in Fig.~\ref{fig1}. The central atom is labeled $1$ and the $N_t$ surrounding atoms are labeled $2,\ldots,N$, where $N=N_t+1$ is the total atom number. In the ideal asymmetric-blockade limit, microwave dressing cancels the target-target interaction while retaining a nonzero control-target interaction. 
Under resonant laser driving with Rabi frequency $\Omega$ and zero detuning, the many-body dynamics of the system are governed by the Hamiltonian~\cite{browaeys2020many}
	\begin{align}\label{eq2}
	H = \frac{\Omega}{2} \sum_{i=1}^{N} \sigma_{i}^{x}  + \sum_{j=2}^N V_{1,j} \, n_{1} n_{j}.
\end{align}
where $\sigma_i^x$ is the Pauli-$x$ operator on atom $i$, $n_i = (\sigma_i^z + \mathbb{I})/2$ projects onto the Rydberg state, and $V_{1,j}$ is the control-target interaction. Residual target-target couplings are not included in Eq.~\eqref{eq2}; will lead to systematic error, which is further analyzed in Sec. \ref{sec:implementation} 
To capture the electric-field dependence of $V_{1,j}$, we consider a three-level F\"{o}rster scheme involving the target Rydberg state $|\alpha\rangle$ and two nearby states $|\alpha'\rangle$ and $|\alpha''\rangle$. When the pair state $|\alpha,\alpha\rangle$ is nearly degenerate with the symmetric combination $|+\rangle = (|\alpha',\alpha''\rangle + |\alpha'',\alpha'\rangle)/\sqrt{2}$~\cite{PhysRevX.2.031011}, the interaction can be described by an effective two-level F\"{o}rster Hamiltonian in the subspace $\{|\alpha,\alpha\rangle, |+\rangle\}$:
\begin{equation}
	V^{(\alpha)}=
	\begin{pmatrix}
		\delta(E) & C_3(E)/R^3 \\
		C_3(E)/R^3 & 0
	\end{pmatrix},
\end{equation}
where $\delta(E)=2\epsilon_{\alpha}(E)-\epsilon_{\alpha'}(E)-\epsilon_{\alpha''}(E)$ is the energy defect and $C_3(E)$ is the resonant dipole-dipole matrix element. The magnitude of the avoided-crossing branch used as the interaction scale is
\begin{equation}
	|V(E,R)|=\frac{\sqrt{\delta(E)^2+4|C_3(E)|^2/R^6}-|\delta(E)|}{2}.
	\label{eq5}
\end{equation}
Far from resonance this expression reduces to the van der Waals scaling $|V|\propto R^{-6}$, while at $\delta(E)=0$ it gives the resonant dipole-dipole scaling $|V|\propto R^{-3}$, leading to a significant enlargement of the Rydberg blockade radius~\cite{PhysRevA.78.060702}.
\begin{figure}
	\centering
	\includegraphics[width=8.86cm,height=3.6cm]{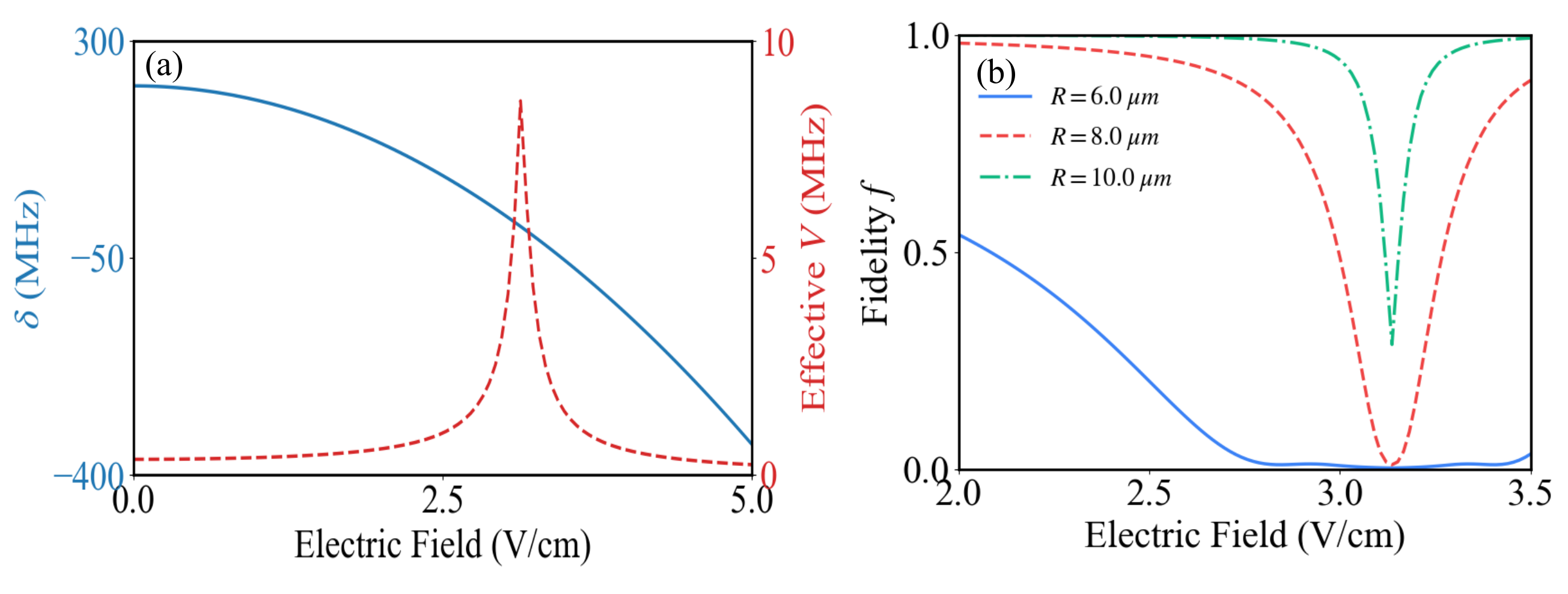}
	\caption{(a) Calculated F\"{o}rster response. The blue curve shows the energy defect $\delta(E)$ between the relevant pair states. The red dashed curve shows the corresponding effective interaction magnitude calculated from Eq.~\eqref{eq5} for the transverse orientation used in the planar-array simulations. (b) Full-excitation probability $f_\tau$ as a function of the electric field for three control-target separations. A sharp dip appears near $E_{\rm res}\simeq3.14~{\rm V/cm}$, where the blockade radius is largest.}
	\label{fig2}
\end{figure}
We evaluate the Stark map with the Alkali Rydberg Calculator (ARC) \cite{SIBALIC2017319}. For the Rb states and dressing parameters listed in Appendix~\ref{app:dressing}, a F\"{o}rster resonance occurs near $E_{\rm res}\simeq3.14~{\rm V/cm}$, as shown in Fig.~\ref{fig2}. The rapid variation of $|V(E,R)|$ around this field is the physical mechanism that converts electric-field changes into population changes.
Fig.~\ref{fig2}(a) shows the calculated energy defect $\delta(E)$ and the resulting effective interaction $|V(E)|$ for the specific pair states $|c\rangle$ and $|t\rangle$ with nearby states $|58S_{1/2}\rangle$ and $|61P_{1/2}\rangle$. A F\"{o}rster resonance occurs near $E_{\mathrm{res}} \approx 3.14\ \mathrm{V/cm}$, where the interaction strength is maximized. This strong field dependence of the interaction forms the physical basis for high-sensitivity electrometry.
	\section{Planar-array electric-field sensing}\label{sec3}
        \begin{figure*}
    	\centering
    	\includegraphics[width=\textwidth,height=5.4cm]{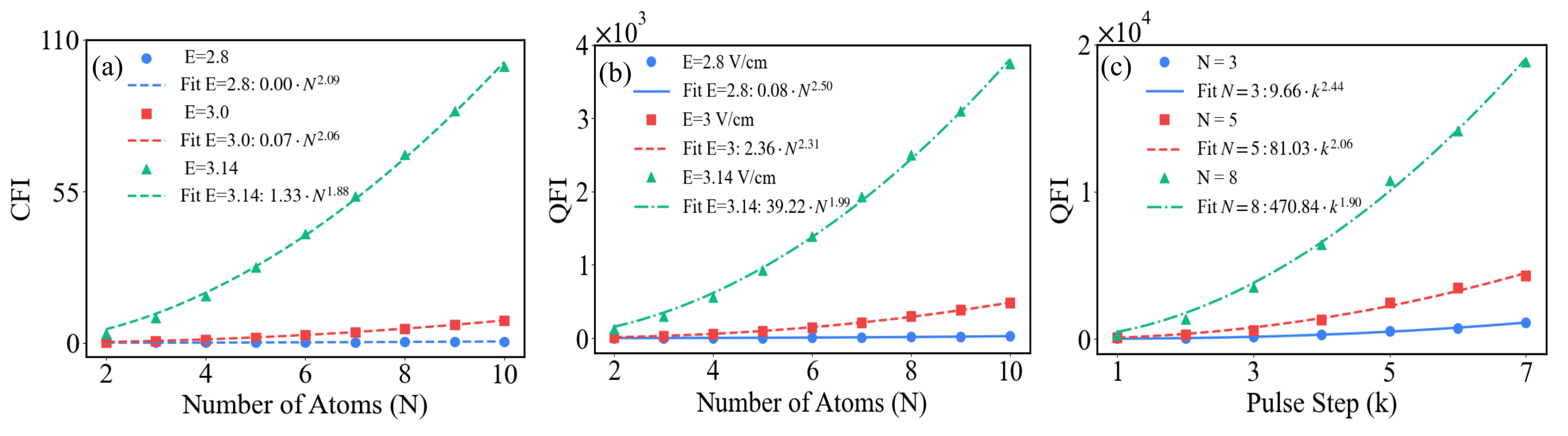}
    	\caption{(a) CFI $F_C$ of the binary full-excitation measurement as a function of total atom number $N$. Power-law fits over the simulated finite-size window yield $F_C\propto N^b$ with $b\simeq2.09$, $2.06$, and $1.88$. (b) QFI versus total atom number for three electric fields. The displayed power-law exponents are fits to the simulated values and should not be interpreted as asymptotic scaling laws. (c) Maximum QFI obtained by RL-optimized composite pulses as a function of pulse number $k$ for several atom numbers. Power-law fits over the displayed finite range give $F_Q\propto k^\gamma$ with $\gamma\simeq2.44$, $2.06$, and $1.90$.}\label{fig3}
    \end{figure*}
  To elucidate the influence of the external electric field $E$ on the sensing response, we investigate the real-time quantum dynamics of the system by numerically integrating the Liouville--von Neumann equation
  \begin{align}
  	\partial_{t}\rho(t) = -i \left[ H(E), \rho(t) \right],
  \end{align}
  where the density operator $\rho(t) = |\Psi(t)\rangle\langle\Psi(t)|$ describes the many-body state and the Hamiltonian $H(E)$ is given by Eq.~(\ref{eq2}), with the control-target interaction term $V_{1,j}(E)$ depending on $E$ through the effective Förster potential. We consider a circular array of $N=4$ target atoms placed on a ring of radius $R = 8~\mu\mathrm{m}$; the central control atom is located at the origin, so that the control--target separation is fixed at $R_{ct}=R$. The system is initialized in the global ground state $|\psi_0\rangle = |0\rangle \otimes |0\rangle^{\otimes N}$ and driven by a uniform laser field with Rabi frequency $\pi$ pulse. The time-domain population dynamics for representative fields are shown in Appendix~\ref{app:dynamics}.
 The experimentally accessible binary signal is the probability that all atoms are found in the Rydberg state at the end of the sequence,
 \begin{align}
 	f_\tau(E)=\left|\langle 1_1,\ldots,1_N|\psi(\tau;E)\rangle\right|^2 .
 	\label{eq3}
 \end{align}
Fig.~\ref{fig2}(b) shows the calculated $f_{\tau}$ as a function of the electric field $E$ for the circular array with $N = 5$ (one control plus four targets) and $R = 6.0, 8.0, 10.0\mu\mathrm{m}$. When $E$ is far from the F\"{o}rster resonance ($E_{\mathrm{res}} \approx 3.14\ \mathrm{V/cm}$), the interatomic interaction is weak, the blockade radius $R_b$ is smaller than the atomic spacing $R$, and the target atoms can be excited nearly independently; consequently, $f_{\tau}$ approaches unity. As $E$ approaches $E_{\mathrm{res}}$, $R_b$ increases dramatically, and the excitation of the central control atom blocks the excitation of all target atoms, causing $f_{\tau}$ to drop sharply to nearly zero, forming a distinct resonance dip. After passing through the resonance, $R_b$ decreases again and $f_{\tau}$ recovers. This strong field dependence of the excitation fidelity provides the basis for high-sensitivity electrometry.

For the final pure state $|\psi(E)\rangle$, the quantum Fisher information (QFI) with respect to the electric field is evaluated by the finite-difference expression \cite{PhysRevLett.134.120803}:
  \begin{align}
  	F_Q(E) = \frac{8\left[1 - |\langle \Psi(E) | \Psi(E+\delta E)\rangle|\right]}{(\delta E)^2},
  	\label{eq:qfi_formula}
  \end{align}
  where $\delta E$ is a small perturbation in the electric field. This expression quantifies the sensitivity of the quantum state to infinitesimal variations in $E$ by measuring the overlap between neighboring states in parameter space. The QFI sets the ultimate lower bound for the uncertainty of any unbiased estimator through the quantum Cramér--Rao inequality \cite{PhysRevLett.72.3439}: $Var(E)  \geq 1/{M F_Q(E)}$, where $M$ is the number of independent measurements.
  While the QFI represents the theoretical optimum over all possible measurements, in practice one must implement a specific measurement protocol. For the fixed binary measurement in Eq.~\eqref{eq3}, the corresponding classical Fisher information (CFI) is
  \begin{equation}
  	F_C(E) = \sum_x \frac{1}{p(x|E)} \left( \frac{\partial p(x|E)}{\partial E} \right)^2.
  \end{equation}
 In general, the classical Fisher information for any specific measurement cannot exceed the quantum Fisher information: $F_C(E) \le F_Q(E)$. Equality holds only when the measurement is optimal, i.e., when it projects onto the eigenbasis of the symmetric logarithmic derivative. Hence, the gap between $F_Q(E)$ and $F_C(E)$ quantifies the room for improvement in measurement design beyond the simple binary readout.

  Fig.~\ref{fig3}(a) shows the scaling of the classical Fisher information ${F}_C$ with respect to the atom number $N$ for different electric fields under the fixed measurement protocol using $f_\tau$ as the observable. For the resonant field $E = E_{\mathrm{res}} = 3.14~\mathrm{V/cm}$, ${F}_C$ scales as ${F}_C \propto N^{b}$ with $b \simeq 1.88$ in the range $N \leq 10$, which is significantly faster than the standard quantum limit ($b=1$). Even for off-resonant fields ($E = 2.8$ and $3.0~\mathrm{V/cm}$), superlinear scalings with $b \simeq 2.09$ and $2.06$ are observed. These results demonstrate that even under a simple, fixed measurement strategy, the many-body entanglement generated by the asymmetric blockade yields a near-quadratic enhancement in sensitivity with increasing atom number, highlighting the intrinsic advantage of our Rydberg array architecture. 
Fig.~\ref{fig3}(b) shows the scaling of $F_Q$ with respect to the atom number $N$ for different electric fields. For the resonant field $E = E_{\mathrm{res}} = 3.14\ \mathrm{V/cm}$, the QFI scales as $F_Q \simeq N^{\alpha}$ with $\alpha \simeq 1.99$ in the range $N \leq 10$, which is significantly faster than the linear scaling expected for uncorrelated probes. Even for off-resonant fields ($E = 2.8$ and $3.0\ \mathrm{V/cm}$), superlinear scalings with $\alpha \simeq 2.50$ and $2.31$ are observed. These results indicate that within the simulated parameter range, increasing the number of sensing atoms provides a more-than-linear payoff in information-extraction capability.

\section{RL-optimized composite pulse sequences}\label{sec:rl}

The single-pulse response can be improved by optimizing the time-domain control. We divide the interrogation into $k$ resonant pulses and allow the phase of the $m$th pulse to be selected from
\begin{equation}
	\mathcal{A}=\{0,\pi/2,\pi,3\pi/2\}.
\end{equation}
During a pulse with phase $\phi_m$, the driving term is
\begin{equation}
	H_{\rm drive}(\phi_m)=\frac{\Omega}{2}\sum_i
	\left(\cos\phi_m\,\sigma_i^x+\sin\phi_m\,\sigma_i^y\right),
\end{equation}
while the interaction term remains that of Eq.~\eqref{eq2}. The RL agent searches over phase sequences to maximize the QFI of the final state. 

    The RL optimization is formulated as a Markov decision process. The state $s_t$ records the phase sequence constructed up to step $t$, and the action $a_t$ selects the next phase from $\mathcal{A}=\{0,\pi/2,\pi,3\pi/2\}$. A fully connected neural network with two hidden layers of 128 and 64 neurons approximates the action-value function. The network is trained with the Adam optimizer using learning rate $5\times10^{-4}$. Exploration is controlled by an $\epsilon$-greedy policy whose value decays from 1.0 to 0.05. 
    An elite replay memory stores the trajectory with the highest QFI found during training. At the end of each episode, the current trajectory is updated with reward based on its QFI, and the elite trajectory is replayed with amplified reward $R_{\rm elite}=(r_{\max}/\zeta)^2$, where $\zeta=40N$. Training uses 800--1000 episodes for the $K=7$ pulse-search examples.	
    The optimization process is detailed in Algorithm 1. Specifically, the agent interacts with the Rydberg environment by selecting a phase index $a_t$ from the set $\mathcal{A}$ at each step $t$, which transitions the state $s_t$ to $s_{t+1}$ by encoding the action into the sequence vector. Upon completing a sequence of length $K$, the QFI ($r_e$) is calculated to generate the reward $R_{curr}$. A distinctive feature of our approach is the dual-stage update: the network $Q(s, a; w)$ is first updated using the current trajectory $\tau_{curr}$, and subsequently reinforced using the elite trajectory $B_{elite}$ with an amplified reward $R_{elite} = (r_{max} / \zeta)^2$ to consolidate the optimal control strategy.
    \DontPrintSemicolon
    \SetNlSty{textnormal}{}{}
    \SetNlSkip{1em}
    \SetInd{0.5em}{1em}	
    \begin{algorithm}[t]
    	\small
    	\caption{Reinforcement learning with elite replay}
    	\textbf{Input:} atom number $N$, sequence length $K$, phase set $\mathcal{A}$\;
    	\textbf{Output:} optimized pulse sequence $\tau^*$\;
    	Initialize Q-network $Q(s,a;w)$ with random weights\;
    	Initialize $r_{\max}\leftarrow0$ and $B_{\rm elite}\leftarrow\emptyset$\;
    	\For{$e=1$ \KwTo $N_{\rm eps}$}{
    		Initialize $s_0$ and current trajectory $\tau_{\rm curr}\leftarrow\emptyset$\;
    		\For{$t=0$ \KwTo $K-1$}{
    			Select $a_t$ via $\epsilon$-greedy policy\;
    			Apply phase $\phi_t=\mathcal{A}[a_t]$ and observe $s_{t+1}$\;
    			Record transition $(s_t,a_t)$ in $\tau_{\rm curr}$\;
    		}
    		Evaluate $r_e\leftarrow F_Q(\tau_{\rm curr},N)$\;
    		\If{$r_e>r_{\max}$}{
    			$r_{\max}\leftarrow r_e$; $B_{\rm elite}\leftarrow\tau_{\rm curr}$\;
    		}
    		Update $w$ using the current trajectory and the elite trajectory\;
    		$\epsilon\leftarrow\max(\epsilon\alpha,\epsilon_{\min})$\;
    	}
    	\Return{$B_{\rm elite}$}\;
    \end{algorithm}

    Figure~\ref{fig3}(c) shows that the optimized phase sequences increase $F_Q$ with pulse number. The fitted exponents are close to quadratic in the present finite-depth regime. Physically, the phase sequence controls the interference between different many-body pathways and steers the final state toward regions of Hilbert space with a larger electric-field derivative. 
    These results demonstrate that RL-discovered pulse sequences construct constructive quantum interference pathways in the time domain, steering the many-body state evolution toward regions of the Hilbert space that are highly sensitive to parameter variations. R-L provides a method for automatically searching for discrete phase sequences. Under the conditions of a fixed pulse depth and a fixed interrogation time, the optimized sequence achieves nearly quadratic fitting, approaching the Heisenberg limit.
  \section{Vectorial Electric Field Sensing with a Spherical Array}\label{sec4} 

	\begin{figure}[htbp]
		\centering
		\includegraphics[width=4.4cm,height=4cm]{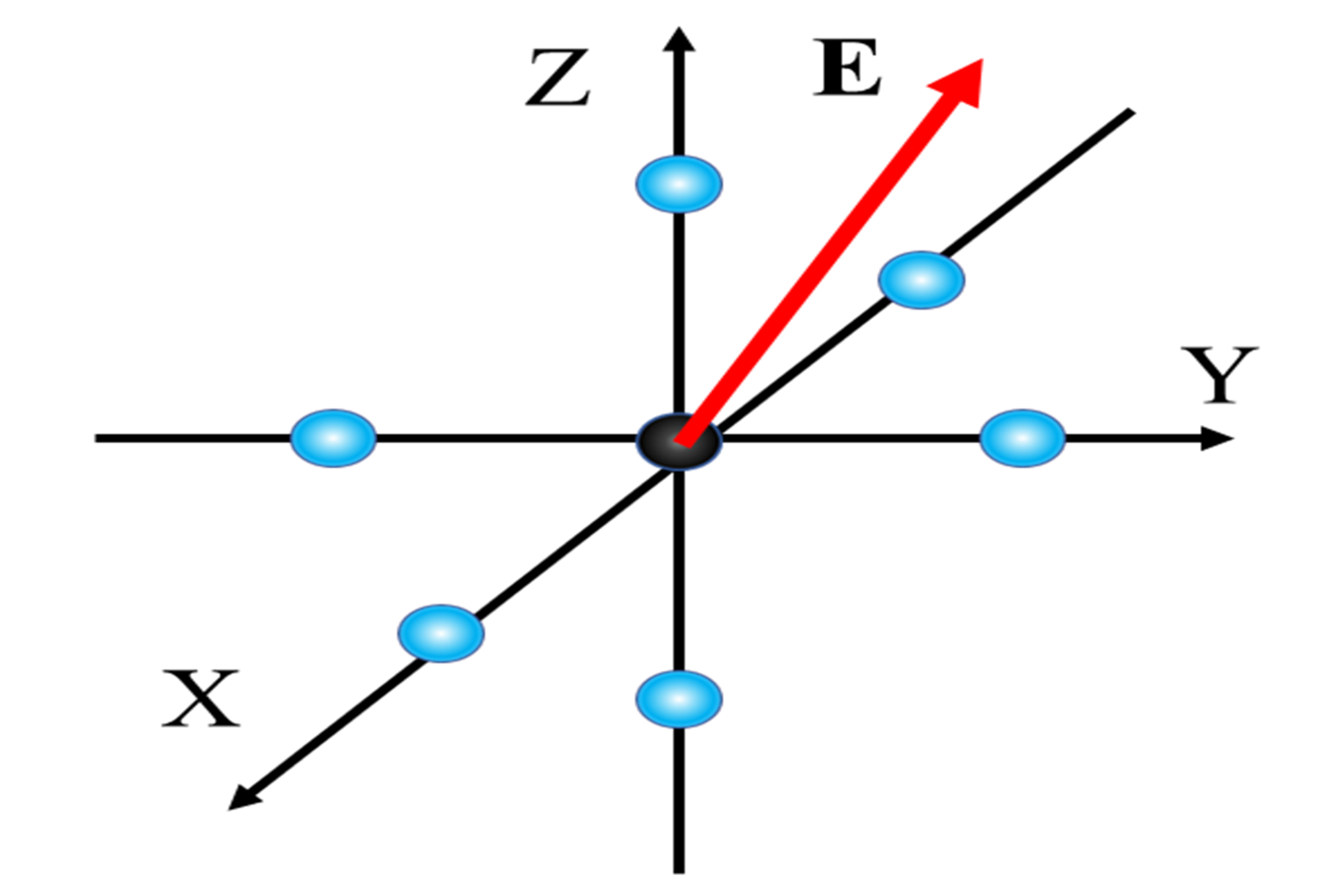}
		\caption{Minimal spherical array for vector sensing. Six target atoms are placed on the Cartesian axes at $(\pm R,0,0)$, $(0,\pm R,0)$, and $(0,0,\pm R)$ around a central control atom.}\label{fig8}
	\end{figure}

	The previous sections focused on sensing the magnitude of an electric field along a fixed direction. Here we extend the scheme to reconstruct the direction of a three-dimensional vector electric field $\mathbf{E} = (E_x, E_y, E_z)$. We propose a minimal spherical array: a single control atom at the center surrounded by six target atoms placed on the Cartesian axes at positions $(\pm R, 0, 0)$, $(0, \pm R, 0)$, and $(0, 0, \pm R)$, as illustrated in Fig.~\ref{fig8}. Through microwave dressing, the interactions between target atoms are completely suppressed ($V_{tt}=0$), while the dipole-dipole interaction between the central atom and each target atom is preserved. For a target atom at position $\mathbf{r}_i$, the interaction strength is given by
	\begin{equation}
	V_{i} = \frac{C_3}{R^3} \left( 1 - 3\cos^2\theta_i \right),
	\end{equation}
	where $\theta_i$ is the angle between the electric field direction $\hat{\mathbf{E}}$ and the displacement vector $\mathbf{r}_i$ of the $i$th target atom. Let the unknown electric field be $\mathbf{E} = (E_x, E_y, E_z)$ with direction cosines $(n_x, n_y, n_z) = (E_x/E_0, E_y/E_0, E_z/E_0)$. For this symmetric configuration, the cosine of the angle for each axial pair of atoms simplifies to:
\begin{subequations}
		\begin{align}
		\cos\theta_{x\pm} &= \pm n_x, \\
		\cos\theta_{y\pm} &= \pm n_y, \\
		\cos\theta_{z\pm} &= \pm n_z.
	\end{align}
\end{subequations}
		\begin{figure}[htbp]
	\centering
	\includegraphics[width=8.6cm,height=3.6cm]{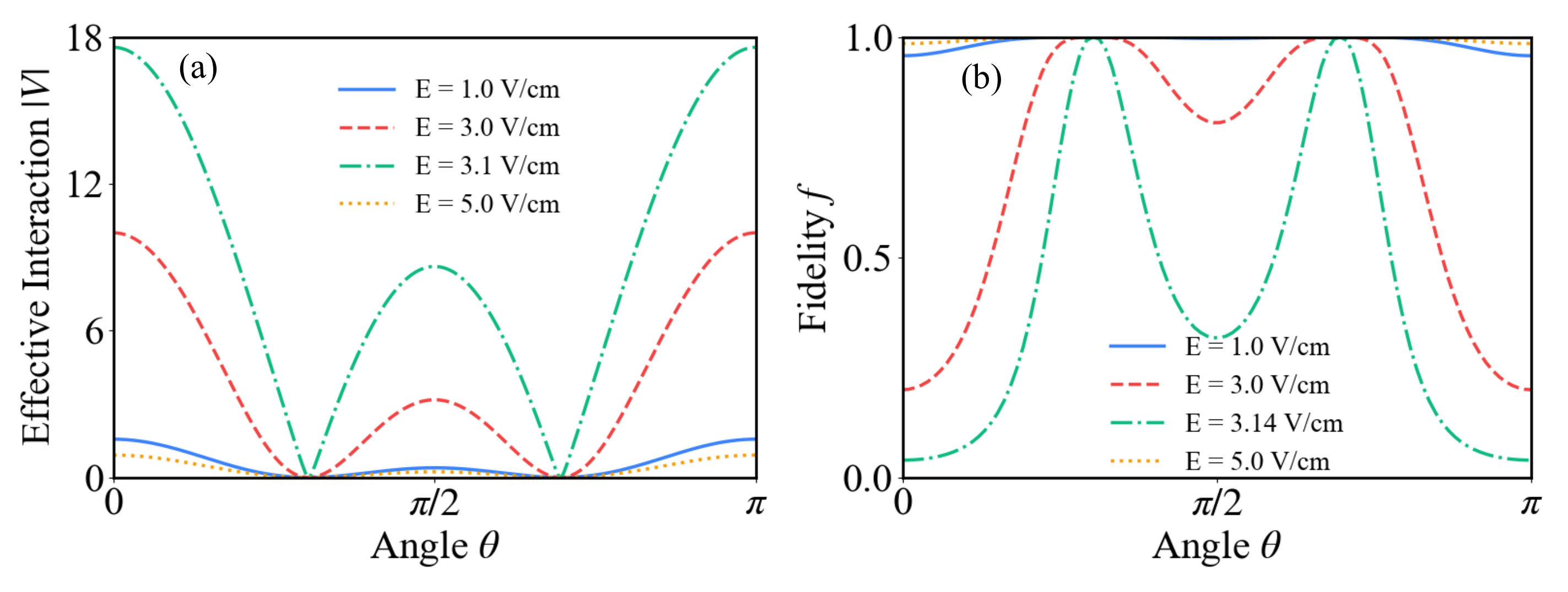}
	\caption{(a) Angular dependence of the effective interaction magnitude for several electric fields. The node occurs at the magic angle $\theta_m=\arccos(1/\sqrt{3})\simeq0.955$ rad. Near the F\"{o}rster resonance, the interaction is enhanced without changing the basic dipolar angular dependence. (b) Axial-pair excitation signal as a function of field angle for several electric-field strengths. The steep angular variation near resonance provides the directional sensitivity required for vector-field reconstruction.}\label{fig4}
\end{figure}

	Substituting these into the interaction formula yields the effective axial interaction energies:

	\begin{subequations}
		\begin{align}
			V_x &= \frac{C_3}{R^3}(1 - 3n_x^2), \\
			V_y &= \frac{C_3}{R^3}(1 - 3n_y^2), \\
			V_z &= \frac{C_3}{R^3}(1 - 3n_z^2).
		\end{align}
	\end{subequations}
	Thus, the electric field direction $(n_x, n_y, n_z)$ is encoded into three independent interaction strengths, which further modulate the excitation probabilities of atoms along each axis. The sensing protocol consists of three steps. First, the system is initialized in the global ground state and then driven by a resonant laser field with Rabi frequency $\Omega$. After evolving for a fixed duration $\tau$, the excitation probabilities of the two target atoms on each axis are measured and averaged, yielding the axial population vector $\mathbf{P} = (P_x, P_y, P_z)$. Finally, vector reconstruction is performed via a pre-calibrated mapping $\mathbf{n} = \mathcal{F}(\mathbf{P})$ to infer the field direction $\hat{\mathbf{E}}$, while the field magnitude $E_0$ is independently calibrated through the Förster-resonance dependence of $C_3$. 
	We note that the interaction $V \propto (1 - 3\cos^2\theta)$ depends only on $\cos^2\theta$, leading to a sign ambiguity in each axis. Due to the dependence of the interaction on $\cos^2\theta$, the excitation probabilities are invariant under the transformation $\hat{\mathbf{E}} \to -\hat{\mathbf{E}}$ as well as under reflection with respect to the magic angle $\theta_m$ where $V=0$. This ambiguity is inherent to the dipole-dipole interaction form and cannot be resolved by the symmetric six-atom array alone. However, it can be eliminated by applying a small known bias field $\mathbf{E}_{\text{bias}}$ along a reference direction and comparing the measured responses with and without the bias, as detailed in Appendix~\ref{app:bias}. With this additional step, the full vector direction can be uniquely determined. Thus, the proposed spherical array together with the bias-field calibration enables full vector electrometry.
	
	Fig.~\ref{fig4}(a) shows the angular dependence of the effective Rydberg interaction $|V|$ for different electric field strengths. Near the F\"{o}rster resonance ($E \approx 3.14\ \mathrm{V/cm}$), the interaction strength is significantly enhanced while preserving the $(1 - 3\cos^2\theta)$ angular shape, which amplifies the direction-dependent blockade effect without distorting the angular encoding. Fig.~\ref{fig4}(b) plots the excitation probability of all atoms being excited to the Rydberg state at the end of the evolution for an axial atom pair as a function of the field angle $\theta$ at the F\"{o}rster resonance. The steep variation of the excitation probability with $\theta$ provides the high directional sensitivity required for precise vector reconstruction. 
	Together, these features enable a compact, single-shot vector electrometry protocol that goes beyond conventional scalar or one-dimensional Rydberg sensing. With the bias-field disambiguation step, our approach fully resolves the dipole-dipole angular ambiguity and reconstructs the three-dimensional field direction from a minimal spherical array, providing new insights into the development of integrated quantum sensors with true vector capability.

\section{Implementation considerations}\label{sec:implementation}

The protocol is designed around observables that are standard in neutral-atom Rydberg experiments: state-selective population readout after a controlled pulse sequence. The required calibrations are (i) the Stark-map relation $\delta(E)$ and $C_3(E)$ near the selected F\"{o}rster channel, (ii) the population response $f_\tau(E)$ for the planar array or $\mathbf{P}(\mathbf{E})$ for the spherical array, and (iii) the residual target-target interaction after microwave dressing. The calculations use separations of $R=6$--$10~\mu{\rm m}$ and Rb Rydberg states with parameters specified in Appendix~\ref{app:dressing}, which are compatible with optical-tweezer array length scales.
Several limitations should be kept explicit. The strongest scaling statements are obtained from closed-system simulations over finite atom numbers and finite pulse depths. Decoherence, imperfect state preparation, atom loss, microwave-dressing errors, and finite readout fidelity will reduce the achievable Fisher information.
To evaluate the performance of our scheme under realistic conditions, we incorporate both dissipative and dephasing processes in our simulations. These parameters are set based on the experimental conditions, with dimensionless decay rate $\gamma_{dec} = 0.0005$ and dephasing rate $\gamma_{dep} = 0.001$ (both normalized by the Rabi frequency $\Omega$), corresponding to a Rydberg state lifetime of approximately $300~\mu\text{s}$ \cite{SIBALIC2017319}.
The dynamics of the $N$-atom system is governed by the Lindblad master equation:
\begin{align}
\dot{\rho} = -i[H, \rho] + \sum_{j} \gamma_{dec} \mathcal{D}[\sigma_-^{(j)}]\rho + \sum_{j} \gamma_{dep} \mathcal{D}[\sigma_z^{(j)}]\rho, 
\end{align}
 where $\mathcal{D}[L]\rho = L\rho L^\dagger - \frac{1}{2}\{L^\dagger L, \rho\}$ is the Lindblad dissipator. Here, $\sigma_-^{(j)}$ represents the decay process of the $j$-th atom at a rate $\gamma_{dec}$, while $\sigma_z^{(j)}$ represents the  dephasing process at a rate $\gamma_{dep}$.	 
	 
	 \begin{figure}
	 	\centering
	 	\includegraphics[width=8.6cm,height=6cm]{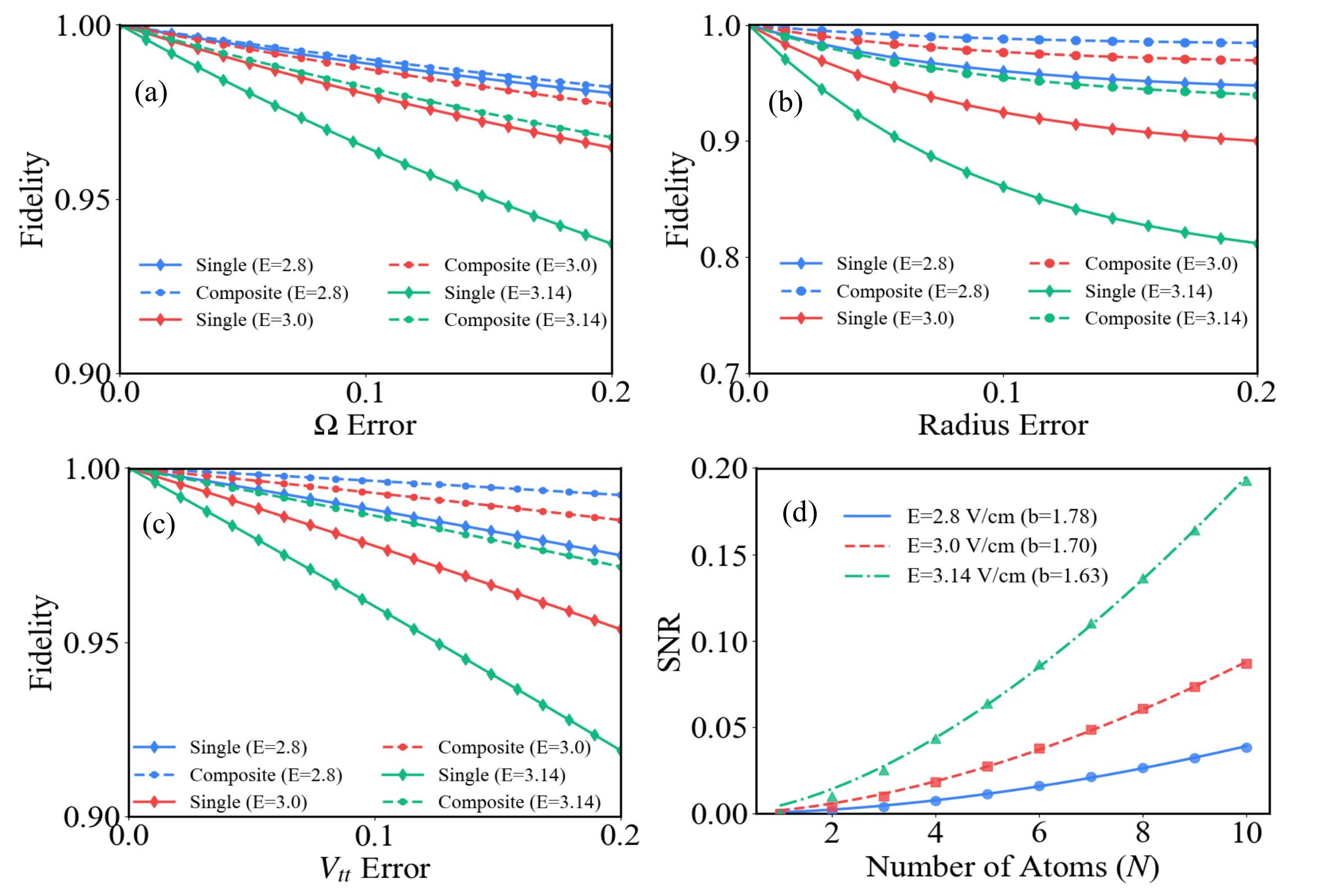}
	 	\caption{(a) Robustness of single-pulse and composite-pulse schemes against Rabi-frequency errors. (b) Robustness of single-pulse and composite-pulse schemes against radial-position errors. (c) Robustness of single-pulse and composite-pulse schemes against residual target-target interaction errors. (d) Signal-to-noise ratio as a function of atom number under projection noise. The fitted exponents are shown in the legend.}\label{fig6}
	 \end{figure}

We define the signal-preservation metric $F_{\rm rel}=1-|F_{\rm ideal}-F_{\rm noisy}|$, where $F_{\rm ideal}$ and $F_{\rm noisy}$ are the ideal and noisy final full excitation state fidelity. Fig.~\ref{fig6}(a), (b) and (c) compare single-pulse and composite-pulse protocols under Rabi-frequency, radius, and residual-interaction errors. The composite pulse is more robust in these numerical tests, consistent with the error-canceling behavior of phase-modulated pulse sequences \cite{levitt1986composite}. 
Under realistic conditions with quantum projection noise, the signal-to-noise ratio (SNR) is computed using the standard formula for binomial projection noise \cite{RevModPhys.89.035002}:
	 \begin{equation}
	 	\text{SNR} = \frac{|P(E + \delta E) - P(E)|}{\sqrt{P(E)(1 - P(E)) / N_{\text{meas}}}},
	 \end{equation}
	 where $P(E)$ is the excitation probability of the fully excited state, and $N_{\text{meas}}$ is the number of independent measurements.	
	 Fig.~\ref{fig6}(d) shows that the SNR increases with atom number in the simulated range. As with the Fisher-information fits, this is a finite-size trend that must be benchmarked experimentally.
	 
Collectively, these implementation-oriented analyses demonstrate that our protocol is not merely a theoretical proposal but a practically viable scheme for Rydberg electrometry. The demonstrated robustness to typical experimental imperfections, combined with the moderate calibration overhead and the compatibility with standard optical-tweezer arrays, establishes a concrete pathway toward realising full vector electric-field sensing with quantum-enhanced sensitivity in the near term.
\section{CONCLUSION}\label{sec5} 
	
We have proposed a Rydberg electric-field sensor based on asymmetric blockade and F\"{o}rster-enhanced control-target interactions. In a planar geometry, the field-dependent blockade produces a sharp full-excitation population response and a finite-size superlinear scaling of Fisher information with atom number. RL-optimized composite pulse phases provide an additional control knob that improves the final-state QFI within the simulated pulse-depth range. In a spherical geometry, axial population measurements enable vector sensing once a calibrated bias-field step is used to remove angular degeneracies. The work therefore provides a concrete application-oriented protocol for Rydberg electrometry, with clear observables, calibration requirements, and robustness checks. Experimental validation of the finite-size scaling and of the vector reconstruction is the next essential step

	\section*{acknowledgements}
	The authors acknowledge the financial support by National Natural Science Foundation of China~(Grants No.~62471001, No.~12475009, No.~12575032, No.~12075001, and No.~12175001), Innovation Program for Quantum
	Science and Technology ~(Grant No. 2582021ZD0301704), Natural Science Research Project in Universities of Anhui Province (Grant No. 2024AH050068), Anhui Provincial Key Research and Development Plan (Grant No. 2022b13020004), Anhui Province Science and Technology Innovation Project (Grant No. 202423r06050004).
	\appendix
	\section{SUPPLEMENTAL MATERIAL A}\label{app:dressing}

Here we provide the explicit microwave dressing configuration used to achieve the asymmetric blockade described in Sec.~\ref{sec2}. The dressing scheme follows Ref.~\cite{PhysRevLett.127.120501}.
We apply two microwave fields coupling one \(s\)-state (\(L=0\)) to two \(p\)-states (\(L=1\)) with different principal quantum numbers. The microwave Hamiltonian in the rotating frame reads
	\begin{equation}
		\begin{aligned}
			H_{\text{mw}} =& -\Delta_0 |p_0\rangle\langle p_0| + \Omega_0 |s\rangle\langle p_0| + \Omega_0^* |p_0\rangle\langle s| \\
			&- \Delta_+ |p_+\rangle\langle p_+| + \Omega_+ |s\rangle\langle p_+| + \Omega_+^* |p_+\rangle\langle s|,
		\end{aligned}
	\end{equation}
	where \(\Delta_{0/+} = \nu_{0/+}-\omega_{0/+}\) denote detunings and \(\Omega_{0/+}\) Rabi frequencies. 
	\begin{equation}
		\begin{aligned}
			V_{dd}^{(i,j)} =& \frac{1 - 3 \cos^2 \theta_{ij}}{R_{ij}^3} \bigl( \mu_0^2 |s_i p_{j,0}\rangle \langle p_{i,0} s_j| \\
			&- \mu_+^2 / 2 |s_i p_{j,+}\rangle \langle p_{i,+} s_j| \bigr) + \text{H.c.},
		\end{aligned}
	\end{equation}
	where $ R_{ij} $ is the distance between atoms $i$ and $ j $, $\theta_{ij}$ is the angle the displacement vector makes with the quantization axis, and 
	$
	\mu_0 = \langle p_0 | d_0 | s \rangle, \quad \mu_+ = \langle p_+ | d_+ | s \rangle
	$ 
	are the transition dipole moments. Here \( d_p = \hat{\mathbf{e}}_p \cdot \mathbf{d} \) denotes the component of the dipole operator \( \mathbf{d} \) along the polarization vector \( \hat{\mathbf{e}}_p \), with $\hat{\mathbf{e}}_0 = \hat{\mathbf{z}}, \qquad \hat{\mathbf{e}}_{\pm} = \mp {\hat{\mathbf{x}} \pm i\hat{\mathbf{y}}}/{\sqrt{2}} .$
	By dressing the central atom into a state \(\ket{c}\propto \ket{s}+c_0\ket{p_0}+c_+\ket{p_+}\) and each target atom into \(\ket{t}\propto \ket{s}+t_0\ket{p_0}+t_+\ket{p_+}\), and choosing coefficients such that \(|t_+|^2 = 2M^2|t_0|^2\) with \(M=\mu_0/\mu_+\), the target-target dipole–dipole interaction vanishes: \(V_{tt}=0\). In contrast, the control–target interaction \(V_{ct}\) remains strong and scales as \(1/R^3\). This asymmetric blockade enables clean sensing of electric fields via the excitation dynamics of the target atoms, without interference from target–target couplings.
	In this paper, we consider the specific microwave-dressed states of \(^{87}\text{Rb}\): 
	\(|s\rangle = |n=60, L=0, J=1/2, m_J=1/2\rangle\), 
	\(|p_0\rangle = |n=60, L=1, J=1/2, m_J=1/2\rangle\), 
	and \(|p_+\rangle = |n=59, L=1, J=1/2, m_J=-1/2\rangle\). 
	The microwave fields couple \(|s\rangle\) to \(|p_0\rangle\) and \(|p_+\rangle\) with Rabi frequencies \(\Omega_0\) and \(\Omega_+\), respectively. And in here, the dressing parameters are chosen as
	$
	\Omega_0/2\pi = -265\;\text{MHz},\quad
	\Delta_0/2\pi = -223\;\text{MHz},\quad
	\Omega_+/2\pi = 176\;\text{MHz},\quad
	\Delta_+/2\pi = 200\;\text{MHz}.
	$ 
	Through this dressing, we construct the control state \(|c\rangle\) and the target state \(|t\rangle\) as superpositions of \(|s\rangle\), \(|p_0\rangle\), and \(|p_+\rangle\), enabling the complete suppression of target-target interactions while preserving a strong control-target interaction. 
\begin{figure}
	\centering
	\includegraphics[width=8.6cm,height=6.6cm]{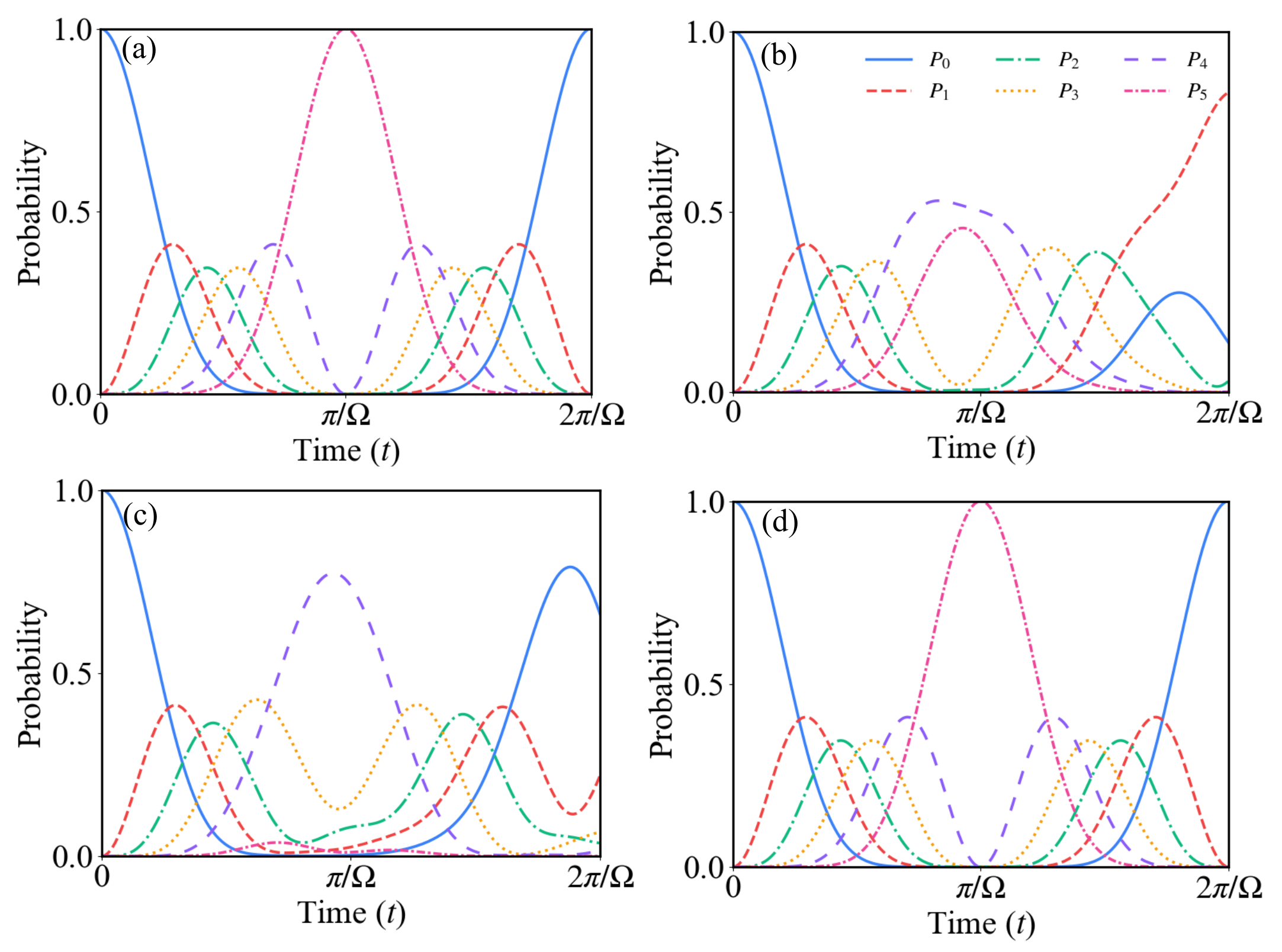}
	\caption{Population dynamics of the five-atom system over one Rabi period for representative electric fields. Close to the F\"{o}rster resonance, the enhanced blockade suppresses the fully excited population.}
	\label{fig5}
\end{figure}
	\section{SUPPLEMENTAL MATERIAL B}\label{app:dynamics}
	
	Fig.~\ref{fig5} shows the changes in the dynamics of the system under four typical electric field intensities. When the electric field is far from the Förster resonance (with electric field intensities of $E = 0~\mathrm{V/cm}$ and $E = 5.0~\mathrm{V/cm}$), the interaction is in the van der Waals region, and the shielding radius is smaller than the atomic distance. The target atoms almost independently undergo excitation - this leads to oscillations in the system's population between the ground state and the fully excited state. As the field approaches the resonance ($E = 2.0~\mathrm{V/cm}$), the energy defect $\delta(E)$ diminishes and the interaction strengthens, crossing over toward the resonant dipole-dipole ($R^{-3}$) scaling. Consequently, the blockade radius expands and suppresses synchronous excitation of the target atoms, reducing the oscillation amplitude. Precisely at the Förster resonance $E_{\mathrm{res}} \approx 3.14~\mathrm{V/cm}$ (where $\delta(E)=0$), the blockade radius reaches its maximum, completely preventing any target atom from being excited while the central atom is in the Rydberg state; during this period, the population in the fully excited state remains close to zero throughout.

	\begin{figure}[htbp]
		\centering
 		\includegraphics[width=5.6cm,height=4cm]{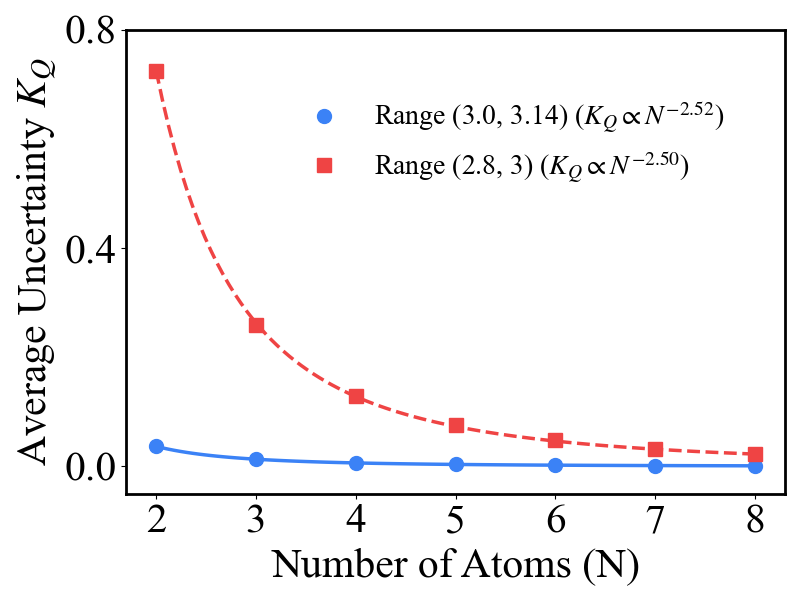}
		\caption{The per-measurement uncertainty $K_Q$ as a function of $N$ for two electric-field ranges: $[2.8, 3.0]\ \text{V/cm}$ (blue) and $[3.0, 3.14]\ \text{V/cm}$ (red). The data follow $K_Q \propto N^{-\beta}$ with $\beta \approx 2.50$ and $2.52$, respectively, demonstrating a significant quantum advantage in precision with increasing array size.}\label{fig7}
	\end{figure}
	
	Correspondingly, the theoretical lower bound for the estimation uncertainty per measurement, defined as 
	\begin{equation}
		K_Q = \int_{\Delta E} f(E) \,  \frac{1}{F_Q(E)} \, dE,
		\label{eq10}
	\end{equation}
	where $ f(E) $ denotes the prior probability distribution of $ E $ over the interval $ \Delta E $, which we assume to be uniform in the following analysis. We then compute the average uncertainty in estimating $ E $ and observe that it decreases rapidly with increasing $ N $, as illustrated in Fig.~\ref{fig7}. The scaling obeys a power-law $ K_Q \propto N^{-\beta} $, with $ \beta \approx 2.50 $ and $ 2.52 $ for the electric-field ranges $ [2.8, 3.0] $ V/cm and $ [3.0, 3.14] $ V/cm, respectively. Notably, the uncertainty is lower in the range closer to the Förster resonance ($ [3.0, 3.14] $ V/cm), indicating enhanced sensitivity near resonance.

\section{SUPPLEMENTAL MATERIAL C}\label{app:bias}

As discussed in Sec.~V, the interaction $|V| \propto |1-3\cos^2\theta|$ depends only on $\cos^2\theta$, leading to sign ambiguity ($\theta \to \pi - \theta$) and magic-angle reflection ambiguity ($\theta \to 2\theta_m - \theta$), where $\theta_m = \arccos(1/\sqrt{3})$ is the magic angle. Both transformations yield identical $|V|$ in the unbiased measurement.
From the first measurement (without bias), we obtain $|V_0|$ for each axis. Solving $|V_0| = \frac{C}{R^3}|1-3\cos^2\theta|$ yields two possible angles per axis: $\theta_1$ and $\theta_2$, where $\theta_1 < \theta_m$ and $\theta_2 > \theta_m$, or vice versa. Both angles give the same $|V_0|$, so the unbiased measurement alone cannot distinguish between them.
To resolve this ambiguity, we apply a small known bias field $\mathbf{E}_{\text{bias}} = E_b \hat{\mathbf{z}}$. The total field becomes $\mathbf{E}_{\text{total}} = \mathbf{E} + E_b \hat{\mathbf{z}}$. We then perform a second measurement to obtain $V_{\text{bias}}^{(x)}$, $V_{\text{bias}}^{(y)}$, and $V_{\text{bias}}^{(z)}$ simultaneously from the three axial pairs.
The key is that after applying the bias, the direction cosines for all three axes are modified. For the $z$-axis,
\begin{align}
n_z' = \frac{E_0 \cos\theta_z + E_b}{\sqrt{E_0^2 + E_b^2 + 2E_0E_b \cos\theta_z}},
\end{align}
where $\cos\theta_z = n_z$ is the $z$-direction cosine of the original field. For the $x$-axis, the original direction cosine $n_x$ is also renormalized:
\begin{subequations}
	\begin{align}
		n_x' &= \frac{E_0 n_x}{\sqrt{E_0^2 + E_b^2 + 2E_0E_b n_z}}.\\
		n_y'&=\frac{E_0n_y}{\sqrt{E_0^2+E_b^2+2E_0E_bn_z}}.
	\end{align}
\end{subequations}
The second measurement gives
\begin{subequations}
	\begin{align}
		V_{\text{bias}}^{(z)} = \frac{C}{R^3}\left(1 - 3n_z'^2\right),\\
		V_{\text{bias}}^{(x)} = \frac{C}{R^3}\left(1 - 3n_x'^2\right),\\
		V_{\text{bias}}^{(y)} = \frac{C}{R^3}\left(1 - 3n_y'^2\right).
	\end{align}
\end{subequations}
For each candidate $\theta_z$ (i.e., each sign and each side of the magic angle), we compute the theoretical $V_{\text{bias}}^{(x)}$, $V_{\text{bias}}^{(y)}$, and $V_{\text{bias}}^{(z)}$ using the above formulas, with $n_x$ and $n_y$ determined from the unbiased measurement (up to sign). By comparing the measured values with the theoretical predictions, the correct candidate is uniquely identified. A single bias field along $+\hat{z}$ thus suffices to resolve all ambiguities for all three components simultaneously.
		
		\bibliography{REV}

@book{hitchcock2004radio,
	author = {Hitchcock, R.T.},
	isbn = {9780471125471},
	title = {Radio-Frequency and Microwave Radiation},
	booktitle = {Patty's Toxicology},
	chapter = {101},
	pages = {133-168},
	doi = {10.1002/0471435139.tox101.pub2},
	year = {2012}
}

@ARTICLE{3601,
	author = {Kraus, J.D.},
	journal = {IEEE Trans. Microwave Theory Tech.},
	title = {Heinrich Hertz-theorist and experimenter},
	year = {1988},
	volume = {36},
	number = {5},
	pages = {824-829},
	doi = {10.1109/22.3601}
}

@ARTICLE{1589516,
	author = {Minasian, R.A.},
	journal = {IEEE Trans. Microwave Theory Tech.},
	title = {Photonic signal processing of microwave signals},
	year = {2006},
	volume = {54},
	number = {2},
	pages = {832-846},
	doi = {10.1109/TMTT.2005.863060}
}

@article{RevModPhys.89.035002,
	author = {Degen, C. L. and Reinhard, F. and Cappellaro, P.},
	title = {Quantum sensing},
	journal = {Rev. Mod. Phys.},
	volume = {89},
	number = {3},
	pages = {035002},
	year = {2017},
	doi = {10.1103/RevModPhys.89.035002}
}

@article{RevModPhys.82.1155,
	author = {Clerk, A. A. and Devoret, M. H. and Girvin, S. M. and Marquardt, F. and Schoelkopf, R. J.},
	title = {Introduction to quantum noise, measurement, and amplification},
	journal = {Rev. Mod. Phys.},
	volume = {82},
	number = {2},
	pages = {1155--1208},
	year = {2010},
	doi = {10.1103/RevModPhys.82.1155}
}

@article{Adams_2020,
	author = {Adams, C.S. and Pritchard, J.D. and Shaffer, J.P.},
	title = {Rydberg atom quantum technologies},
	journal = {J. Phys. B: At. Mol. Opt. Phys.},
	volume = {53},
	number = {1},
	pages = {012002},
	year = {2019},
	doi = {10.1088/1361-6455/ab52ef}
}

@article{Fan_2015,
	author = {Fan, H.Q. and Kumar, S. and Sedlacek, J. and Kübler, H. and Karimkashi, S. and Shaffer, J.P.},
	title = {Atom based RF electric field sensing},
	journal = {J. Phys. B: At. Mol. Opt. Phys.},
	volume = {48},
	number = {20},
	pages = {202001},
	year = {2015},
	doi = {10.1088/0953-4075/48/20/202001}
}

@misc{kitson2025sensingelectricfieldsrydberg,
	author = {Kitson, P. and {W. J. C.} and Birkl, G. and Amico, L. and Polo, J.},
	title = {Sensing electric fields through Rydberg atom networks},
	year = {2025},
	eprint = {2509.01665},
	archivePrefix = {arXiv},
	primaryClass = {quant-ph}
}

@article{PhysRevLett.85.2208,
	author = {Jaksch, D. and Cirac, J. I. and Zoller, P. and Rolston, S. L. and C\^ot\'e, R. and Lukin, M. D.},
	title = {Fast Quantum Gates for Neutral Atoms},
	journal = {Phys. Rev. Lett.},
	volume = {85},
	number = {10},
	pages = {2208--2211},
	year = {2000},
	doi = {10.1103/PhysRevLett.85.2208}
}

@article{PhysRevLett.87.037901,
	author = {Lukin, M. D. and Fleischhauer, M. and Cote, R. and Duan, L. M. and Jaksch, D. and Cirac, J. I. and Zoller, P.},
	title = {Dipole Blockade and Quantum Information Processing in Mesoscopic Atomic Ensembles},
	journal = {Phys. Rev. Lett.},
	volume = {87},
	number = {3},
	pages = {037901},
	year = {2001},
	doi = {10.1103/PhysRevLett.87.037901}
}

@book{gallagher1994rydberg,
	author = {Gallagher, T.F.},
	title = {Rydberg Atoms},
	publisher = {Cambridge University Press},
	year = {1994},
	doi = {10.1017/CBO9780511524530}
}

@article{PhysRevA.65.063404,
	author = {Anderson, W. R. and Robinson, M. P. and Martin, J. D. D. and Gallagher, T. F.},
	title = {Dephasing of resonant energy transfer in a cold Rydberg gas},
	journal = {Phys. Rev. A},
	volume = {65},
	number = {6},
	pages = {063404},
	year = {2002},
	doi = {10.1103/PhysRevA.65.063404}
}

@article{PhysRevX.2.031011,
	author = {Nipper, J. and Balewski, J. B. and Krupp, A. T. and Hofferberth, S. and L\"ow, R. and Pfau, T.},
	title = {Atomic Pair-State Interferometer: Controlling and Measuring an Interaction-Induced Phase Shift in Rydberg-Atom Pairs},
	journal = {Phys. Rev. X},
	volume = {2},
	number = {3},
	pages = {031011},
	year = {2012},
	doi = {10.1103/PhysRevX.2.031011}
}

@article{facon2016sensitive,
	author = {Facon, A. and Dietsche, E.K. and Grosso, D. and Haroche, S. and Raimond, J.M. and Brune, M. and Gleyzes, S.},
	title = {A sensitive electrometer based on a Rydberg atom in a Schr\"{o}dinger-cat state},
	journal = {Nature},
	volume = {535},
	number = {7611},
	pages = {262--265},
	year = {2016},
	doi = {10.1038/nature18327}
}

@article{PhysRevLett.134.120803,
	author = {Xu, H. and Xiao, T.L. and Huang, J.Z and He, M. and Fan, J.P and Zeng, G.H},
	title = {Toward Heisenberg Limit without Critical Slowing Down via Quantum Reinforcement Learning},
	journal = {Phys. Rev. Lett.},
	volume = {134},
	number = {12},
	pages = {120803},
	year = {2025},
	doi = {10.1103/PhysRevLett.134.120803}
}

@article{PhysRevLett.127.120501,
	author = {Young, J.T. and Bienias, P. and Belyansky, R. and Kaufman, A.M. and Gorshkov, A.V.},
	title = {Asymmetric Blockade and Multiqubit Gates via Dipole-Dipole Interactions},
	journal = {Phys. Rev. Lett.},
	volume = {127},
	number = {12},
	pages = {120501},
	year = {2021},
	doi = {10.1103/PhysRevLett.127.120501}
}

@ARTICLE{6910267,
	author = {Holloway, C.L. and Gordon, J.A. and Jefferts, S. and Schwarzkopf, A. and Anderson, D.A. and Miller, S.A. and Thaicharoen, N. and Raithel, G.},
	journal = {IEEE Trans. Antennas Propag.},
	title = {Broadband Rydberg Atom-Based Electric-Field Probe for SI-Traceable, Self-Calibrated Measurements},
	year = {2014},
	volume = {62},
	number = {12},
	pages = {6169-6182},
	doi = {10.1109/TAP.2014.2360208}
}

@incollection{walker2012entanglement,
	author = {Walker, T. G. and Saffman, M.},
	title = {Chapter 2 - Entanglement of Two Atoms Using Rydberg Blockade},
	booktitle = {Advances in Atomic, Molecular, and Optical Physics},
	volume = {61},
	pages = {81-115},
	publisher = {Academic Press},
	year = {2012},
	doi = {10.1016/B978-0-12-396482-3.00002-8}
}

@article{ding2022enhanced,
	author = {Ding, D.S. and Liu, Z.K. and Shi, B.S. and Guo, G.C. and M{\o}lmer, K. and Adams, C.S.},
	title = {Enhanced metrology at the critical point of a many-body Rydberg atomic system},
	journal = {Nat. Phys.},
	volume = {18},
	number = {12},
	pages = {1447--1452},
	year = {2022},
	doi = {10.1038/s41567-022-01777-8}
}

@article{jing2020atomic,
	author = {Jing, M.Y and Hu, Y. and Ma, J. and Zhang, H. and Zhang, L.J. and Xiao, L.T. and Jia, S.T},
	title = {Atomic superheterodyne receiver based on microwave-dressed Rydberg spectroscopy},
	journal = {Nat. Phys.},
	volume = {16},
	number = {9},
	pages = {911--915},
	year = {2020},
	doi = {10.1038/s41567-020-0918-5}
}

@article{10.1063/5.0069195,
	author = {Prajapati, N. and Robinson, A.K. and Berweger, S. and Simons, M.T. and Artusio-Glimpse, A.B. and Holloway, C. L.},
	title = {Enhancement of electromagnetically induced transparency based Rydberg-atom electrometry through population repumping},
	journal = {Appl. Phys. Lett.},
	volume = {119},
	number = {21},
	pages = {214001},
	year = {2021},
	doi = {10.1063/5.0069195}
}

@article{PhysRevA.104.043103,
	author = {Meyer, D.H. and O'Brien, C. and Fahey, D. P. and Cox, K. C. and Kunz, P. D.},
	title = {Optimal atomic quantum sensing using electromagnetically-induced-transparency readout},
	journal = {Phys. Rev. A},
	volume = {104},
	number = {4},
	pages = {043103},
	year = {2021},
	doi = {10.1103/PhysRevA.104.043103}
}

@article{Prajapati:21,
	author = {Prajapati, N. and Niu, Z.Q. and Novikova, I.},
	title = {Quantum-enhanced two-photon spectroscopy using two-mode squeezed light},
	journal = {Opt. Lett.},
	volume = {46},
	number = {8},
	pages = {1800--1803},
	year = {2021},
	doi = {10.1364/OL.418398}
}

@article{wade2017real,
	author = {Wade, C. G. and Šibalić, N. and de Melo, N. R. and Kondo, J. M. and Adams, C. S. and Weatherill, K. J.},
	title = {Real-time near-field terahertz imaging with atomic optical fluorescence},
	journal = {Nat. Photonics},
	volume = {11},
	number = {1},
	pages = {40--43},
	year = {2016},
	doi = {10.1038/nphoton.2016.214}
}

@article{Pause:24,
	author = {Pause, L. and Sturm, L. and Mittenb\"uhler, M. and Amann, S. and Preuschoff, T. and Sch\"affner, D. and Schlosser, M. and Birkl, G.},
	title = {Supercharged two-dimensional tweezer array with more than 1000 atomic qubits},
	journal = {Optica},
	volume = {11},
	number = {2},
	pages = {222--226},
	year = {2024},
	doi = {10.1364/OPTICA.513551}
}

@article{PhysRevLett.130.180601,
	author = {Schlosser, M. and Tichelmann, S. and Sch\"affner, D. and de Mello, Da. O. and Hambach, M. and Sch\"utz, J. and Birkl, G.},
	title = {Scalable Multilayer Architecture of Assembled Single-Atom Qubit Arrays in a Three-Dimensional Talbot Tweezer Lattice},
	journal = {Phys. Rev. Lett.},
	volume = {130},
	number = {18},
	pages = {180601},
	year = {2023},
	doi = {10.1103/PhysRevLett.130.180601}
}

@article{browaeys2020many,
	author = {Browaeys, Antoine and Lahaye, Thierry},
	title = {Many-body physics with individually controlled Rydberg atoms},
	journal = {Nat. Phys.},
	volume = {16},
	number = {2},
	pages = {132--142},
	year = {2020},
	doi = {10.1038/s41567-019-0733-z}
}

@article{PhysRevA.78.060702,
	author = {Reinhard, A. and Younge, K. C. and Raithel, G.},
	title = {Effect of F\"orster resonances on the excitation statistics of many-body Rydberg systems},
	journal = {Phys. Rev. A},
	volume = {78},
	number = {6},
	pages = {060702},
	year = {2008},
	doi = {10.1103/PhysRevA.78.060702}
}

@article{SIBALIC2017319,
	author = {Šibalić, N. and Pritchard, J.D. and Adams, C.S. and Weatherill, K.J.},
	title = {ARC: An open-source library for calculating properties of alkali Rydberg atoms},
	journal = {Comput. Phys. Commun.},
	volume = {220},
	pages = {319-331},
	year = {2017},
	doi = {10.1016/j.cpc.2017.06.015}
}

@article{levitt1986composite,
	author = {Levitt, M.H.},
	title = {Composite pulses},
	journal = {Prog. Nucl. Magn. Reson. Spectrosc.},
	volume = {18},
	number = {2},
	pages = {61-122},
	year = {1986},
	doi = {10.1016/0079-6565(86)80005-X}
}

@article{PhysRevLett.72.3439,
	author = {Braunstein, Samuel L. and Caves, Carlton M.},
	title = {Statistical distance and the geometry of quantum states},
	journal = {Phys. Rev. Lett.},
	volume = {72},
	number = {22},
	pages = {3439--3443},
	year = {1994},
	doi = {10.1103/PhysRevLett.72.3439}
}

@article{Wu2025,
	author = {Wu, J. and Wu, J.L. and Guo, F.Q. and Liu, B.B. and Su, S.L. and Song, X.K. and Ye, L. and Wang, D.},
	title = {Quantum computation via Floquet tailored Rydberg interactions},
	journal = {npj Quantum Inf.},
	volume = {11},
	number = {1},
	year = {2025},
	doi = {10.1038/s41534-025-01068-z}
}

@article{Anand2024,
	author = {Anand, S. and Bradley, C.E. and White, R. and Ramesh, V. and Singh, K. and Bernien, H.},
	title = {A dual-species Rydberg array},
	journal = {Nat. Phys.},
	volume = {20},
	number = {11},
	pages = {1744--1750},
	year = {2024},
	doi = {10.1038/s41567-024-02638-2}
}

@article{PhysRevA.109.012619,
	author = {Jin, Z.Y. and Jing, J.},
	title = {Geometric quantum gates via dark paths in Rydberg atoms},
	journal = {Phys. Rev. A},
	volume = {109},
	number = {1},
	pages = {012619},
	year = {2024},
	doi = {10.1103/PhysRevA.109.012619}
}

@article{Wu:2023axc,
	author = {Wu, C.E. and Kirova, T. and Auzins, M. and Chen, Y.H.},
	title = {Rydberg-Rydberg interaction strengths and dipole blockade radii in the presence of F\"{o}rster resonances},
	journal = {Opt. Express},
	volume = {31},
	number = {22},
	pages = {37094},
	year = {2023},
	doi = {10.1364/OE.502183}
}

@article{Urban2009,
	author = {Urban, E. and Johnson, T. A. and Henage, T. and Isenhower, L. and Yavuz, D. D. and Walker, T. G. and Saffman, M.},
	title = {Observation of Rydberg blockade between two atoms},
	journal = {Nat. Phys.},
	volume = {5},
	number = {2},
	pages = {110--114},
	year = {2009},
	doi = {10.1038/nphys1178}
}

@article{PhysRevApplied.19.044007,
	author = {Su, S.L. and Sun, L.N. and Liu, B. J. and Yan, L.L. and Yung, M.H. and Li, W. and Feng, M.},
	title = {Rabi- and Blockade-Error-Resilient All-Geometric Rydberg Quantum Gates},
	journal = {Phys. Rev. Appl.},
	volume = {19},
	number = {4},
	pages = {044007},
	year = {2023},
	doi = {10.1103/PhysRevApplied.19.044007}
}

@article{PhysRevResearch.2.043130,
	author = {Liu, B.J. and Su, S.L. and Yung, M.H.},
	title = {Nonadiabatic noncyclic geometric quantum computation in Rydberg atoms},
	journal = {Phys. Rev. Res.},
	volume = {2},
	number = {4},
	pages = {043130},
	year = {2020},
	doi = {10.1103/PhysRevResearch.2.043130}
}

@article{PhysRevA.102.062410,
	author = {Guo, F.Q. and Wu, J.L. and Zhu, X.Y. and Jin, Z. and Zeng, Y. and Zhang, S. and Yan, L.L. and Feng, M. and Su, S.L.},
	title = {Complete and nondestructive distinguishment of many-body Rydberg entanglement via robust geometric quantum operations},
	journal = {Phys. Rev. A},
	volume = {102},
	number = {6},
	pages = {062410},
	year = {2020},
	doi = {10.1103/PhysRevA.102.062410}
}

@article{PhysRevLett.117.113601,
	author = {Gullans, M. J. and Thompson, J. D. and Wang, Y. and Liang, Q.Y. and Vuletić, V. and Lukin, M. D. and Gorshkov, A. V.},
	title = {Effective Field Theory for Rydberg Polaritons},
	journal = {Phys. Rev. Lett.},
	volume = {117},
	number = {11},
	pages = {113601},
	year = {2016},
	doi = {10.1103/PhysRevLett.117.113601}
}

@article{PhysRevA.109.022613,
	author = {Song, P.Y. and Wei, J.F. and Xu, P. and Yan, L.L. and Feng, M. and Su, S.L. and Chen, G.},
	title = {Fast realization of high-fidelity nonadiabatic holonomic quantum gates with a time-optimal-control technique in Rydberg atoms},
	journal = {Phys. Rev. A},
	volume = {109},
	number = {2},
	pages = {022613},
	year = {2024},
	doi = {10.1103/PhysRevA.109.022613}
}

@article{PRXQuantum.4.020335,
	author = {Fromonteil, C. and Bluvstein, D. and Pichler, H.},
	title = {Protocols for Rydberg Entangling Gates Featuring Robustness against Quasistatic Errors},
	journal = {PRX Quantum},
	volume = {4},
	number = {2},
	pages = {020335},
	year = {2023},
	doi = {10.1103/PRXQuantum.4.020335}
}

@article{Schlossberger2024,
	author = {Schlossberger, N. and Prajapati, N. and Berweger, S. and Rotunno, A. P. and Artusio-Glimpse, A.B. and Simons, M.T. and Sheikh, A.A. and Norrgard, E.B. and Eckel, S. P. and Holloway, C.L.},
	title = {Rydberg states of alkali atoms in atomic vapour as SI-traceable field probes and communications receivers},
	journal = {Nat. Rev. Phys.},
	volume = {6},
	number = {10},
	pages = {606--620},
	year = {2024},
	doi = {10.1038/s42254-024-00756-7}
}

@article{HanYuLong2026,
	author = {Han, Y.L. and Shan, Z.Y. and Zhang, K. and Sun, J.F. and Zhang, L.H. and Liu, B. and Ding, D.S.},
	title = {Research on High-Sensitivity Sensing Technology of 10 MHz Radio Frequency Electric Field Based on Rydberg Atoms},
	journal = {Acta Phys. Sin.},
	volume = {75},
	number = {13},
	year = {2026},
	doi = {10.7498/aps.75.20260151}
}

@misc{su2026broadbandheterodynemicrowavedetection,
	author = {Su, H.J. and Fang, S.C. and Li, T.A. and Chang, C.H. and Chen, Y.C. and Chen, Y.H.},
	title = {Broadband Heterodyne Microwave Detection using Rydberg Atoms with High Sensitivity},
	year = {2026},
	eprint = {2601.19305},
	archivePrefix = {arXiv},
	primaryClass = {physics.atom-ph}
}

@article{Yan2025,
	author = {Yan, H.M. and Jing, M.Y. and Tong, Y.J. and Yang, W.G. and Zhang, H. and Liu, Z.K. and Xie, J.Y. and Zheng, Y.H. and Xiao, L.T. and Jia, S.T. and Zhang, L.J.},
	title = {Sub-shot-noise Rydberg EIT spectrum},
	journal = {PhotoniX},
	volume = {6},
	number = {1},
	year = {2025},
	doi = {10.1186/s43074-025-00215-1}
}

@article{Wang2026,
	author = {Wang, Y.J. and Zhang, J. and Zhang, Z.Y. and Shao, S.Y. and Li, Q. and Chen, H.C. and Ma, Y. and Han, T.Y. and Wang, Q.F. and Nan, J.D. and Yin, Y.M. and Zhu, D.Y. and Fang, Q.Q. and Yu, C. and Liu, X. and Guo, G.C. and Liu, B. and Zhang, L.H. and Ding, D.S. and Shi, B.S.},
	title = {Quantum enhanced metrology based on flipping trajectory of cold Rydberg gases},
	journal = {Nat. Commun.},
	volume = {17},
	number = {1},
	year = {2026},
	doi = {10.1038/s41467-025-67921-z}
}

@article{Ding2022,
	author = {Ding, D.S. and Liu, Z.K. and Shi, B.S. and Guo, G.C. and Mølmer, K. and Adams, C.S.},
	title = {Enhanced metrology at the critical point of a many-body Rydberg atomic system},
	journal = {Nat. Phys.},
	volume = {18},
	number = {12},
	pages = {1447--1452},
	year = {2022},
	doi = {10.1038/s41567-022-01777-8}
}

@article{PhysRevB.111.144313,
	author = {Yang, F. and Yarloo, H. and Zhang, H.C. and Mølmer, K. and Nielsen, A. E. B.},
	title = {Probing Hilbert space fragmentation with strongly interacting Rydberg atoms},
	journal = {Phys. Rev. B},
	volume = {111},
	number = {14},
	pages = {144313},
	year = {2025},
	doi = {10.1103/PhysRevB.111.144313}
}

@misc{desantis2026realizationcavitycoupledrydbergarray,
	author = {Santis, J.D. and Kovács, B. D. and Öncü, M. and Bouscal, A. and Vasileiadis, D. and Zeiher, J.},
	title = {Realization of a cavity-coupled Rydberg array},
	year = {2026},
	eprint = {2602.12152},
	archivePrefix = {arXiv},
	primaryClass = {quant-ph}
}

@article{PhysRevLett.132.113601,
	author = {Ocola, P. L. and Dimitrova, I. and Grinkemeyer, B. and Guardado-Sanchez, E. and Đorđević, T. and Samutpraphoot, P. and Vuletić, V. and Lukin, M. D.},
	title = {Control and Entanglement of Individual Rydberg Atoms near a Nanoscale Device},
	journal = {Phys. Rev. Lett.},
	volume = {132},
	number = {11},
	pages = {113601},
	year = {2024},
	doi = {10.1103/PhysRevLett.132.113601}
}

@article{5591-xjfr,
	author = {Cooke, L. W. and Czischek, S.},
	title = {Reinforcement learning for optimal control of spin magnetometers},
	journal = {Phys. Rev. A},
	volume = {112},
	number = {6},
	pages = {062603},
	year = {2025},
	doi = {10.1103/5591-xjfr}
}

@article{PhysRevA.109.062609,
	author = {Belliardo, F. and Zoratti, F. and Giovannetti, V.},
	title = {Applications of model-aware reinforcement learning in Bayesian quantum metrology},
	journal = {Phys. Rev. A},
	volume = {109},
	number = {6},
	pages = {062609},
	year = {2024},
	doi = {10.1103/PhysRevA.109.062609}
}

@article{Porotti2023,
	author = {Porotti, R. and Peano, V. and Marquardt, F.},
	title = {Gradient-Ascent Pulse Engineering with Feedback},
	journal = {PRX Quantum},
	volume = {4},
	number = {3},
	year = {2023},
	doi = {10.1103/PRXQuantum.4.030305}
}

@article{6gql-zgkb,
	author = {Xiao, L. and Sarkar, S. and Wang, K.K. and Bayat, A. and Xue, P.},
	title = {Observation of Criticality-Enhanced Quantum Sensing in Nonunitary Quantum Walks},
	journal = {Phys. Rev. Lett.},
	volume = {136},
	number = {6},
	pages = {060802},
	year = {2026},
	doi = {10.1103/6gql-zgkb}
}

@article{PhysRevLett.123.230401,
	author = {Xiao, L. and Wang, K.K. and Zhan, X. and Bian, Z.H. and Kawabata, K. and Ueda, M. and Yi, W. and Xue, P.},
	title = {Observation of Critical Phenomena in Parity-Time-Symmetric Quantum Dynamics},
	journal = {Phys. Rev. Lett.},
	volume = {123},
	number = {23},
	pages = {230401},
	year = {2019},
	doi = {10.1103/PhysRevLett.123.230401}
}

@article{PhysRevLett.121.150503,
	author = {Gray, J. and Banchi, L. and Bayat, A. and Bose, S.},
	title = {Machine-Learning-Assisted Many-Body Entanglement Measurement},
	journal = {Phys. Rev. Lett.},
	volume = {121},
	number = {15},
	pages = {150503},
	year = {2018},
	doi = {10.1103/PhysRevLett.121.150503}
}

@article{Montenegro2025,
	author = {Montenegro, V. and Mukhopadhyay, C. and Yousefjani, R. and Sarkar, S. and Mishra, U. and Paris, M. G.A. and Bayat, A.},
	title = {Review: Quantum metrology and sensing with many-body systems},
	journal = {Phys. Rep.},
	volume = {1134},
	pages = {1--62},
	year = {2025},
	doi = {10.1016/j.physrep.2025.05.005}
}

@article{PhysRevA.100.032104,
	author = {Yang, J. and Pang, S.S. and Zhou, Y.Y. and Jordan, A.N.},
	title = {Optimal measurements for quantum multiparameter estimation with general states},
	journal = {Phys. Rev. A},
	volume = {100},
	number = {3},
	pages = {032104},
	year = {2019},
	doi = {10.1103/PhysRevA.100.032104}
}
\end{document}